\documentclass[prl,twocolumn,floatfix]{revtex4}

      \usepackage{epsf}
      \usepackage{graphicx}
      \usepackage{dcolumn}
      \usepackage{bm}

      \newcommand{\be}{\begin{equation}}
      \newcommand{\ee}{\end{equation}}
      \newcommand{\ba}{\begin{eqnarray}}
      \newcommand{\ea}{\end{eqnarray}}
      \newcommand{\nn}{\nonumber \\}

\begin{document}

      \title{Consistency of the Adiabatic Theorem}

      \author{M.~H.~S.~Amin}
      \affiliation{D-Wave Systems Inc., 100-4401 Still Creek Drive,
      Burnaby, B.C., V5C 6G9, Canada}


\begin{abstract}
The adiabatic theorem provides the basis for the adiabatic model of
quantum computation. Recently the conditions required for the
adiabatic theorem to hold have become a subject of some controversy.
Here we show that the reported violations of the adiabatic theorem
all arise from resonant transitions between energy levels. In the
absence of fast driven oscillations the traditional adiabatic
theorem holds. Implications for adiabatic quantum computation is
discussed.
\end{abstract}
\maketitle

A statement of the {\em traditional} adiabatic theorem
\cite{Ehrenfest,Born,Kato}, as described in most recent
publications, is as follows: Consider a system with a time dependent
Hamiltonian $H(t)$ and a wave function $|\psi(t)\rangle$, which is
the solution of the Schr\"odinger equation ($\hbar=1$)
 \be
 i|\dot \psi(t) \rangle = H(t) |\psi(t)\rangle. \label{SchEq}
 \ee
Let $|E_n(t)\rangle$ be the instantaneous eigenstates of $H(t)$ with
eigenvalues $E_n(t)$. If at an initial time $t=0$ the system starts
in an eigenstate $|E_n(0)\rangle$ of the Hamiltonian $H(0)$, it will
remain in the same instantaneous eigenstate, $|E_n(t)\rangle$, at a
later time $t=T$, as long as the evolution of the Hamiltonian is
slow enough to satisfy
 \be
 \max_{t \in [0,T]} \left|{\langle E_m(t)|\dot E_n(t) \rangle
 \over E_{nm}(t)} \right| \ll 1 \qquad \text{for all } m\neq n,
 \label{Acond}
 \ee
where $E_{nm}(t)\equiv E_n(t)-E_m(t)$. One can easily show that:
$\langle E_m(t)|\dot E_n(t) \rangle = \langle E_m(t)|\dot H| E_n(t)
\rangle/E_{nm}(t)$. The adiabatic theorem has recently gained
renewed attention as it provides the basis for one of the important
schemes for quantum computation, i.e., adiabatic quantum computation
\cite{Farhi,Aharonov}.

Recently, the adiabatic condition (\ref{Acond}) has become a subject
of controversy. It was first shown by Marzlin and Sanders
\cite{Sanders} and then by Tong {\em et al.}~\cite{Tong} that if a
first system with Hamiltonian $H(t)$ follows an adiabatic evolution,
a second system defined by Hamiltonian
 \be
 \bar H(t) = -U^\dagger(t)H(t)U(t), \qquad U(t)\equiv {\cal T}
e^{-i\int_0^t H(t)dt}, \label{barH}
 \ee
cannot have an adiabatic evolution unless
 \be
 |\langle E_n(t)|E_n(0)\rangle| \approx 1, \label{STcond}
 \ee
even if both systems satisfy the same condition (\ref{Acond}). Here,
${\cal T}$ denotes time ordering operator. Recently, the validity of
the adiabatic theorem was experimentally examined \cite{Du}, and
(\ref{Acond}) was reported to be neither sufficient nor necessary
condition for adiabaticity.

These inconsistencies have created debates among researchers
\cite{Duki,Ma,Marzlin,WuYang} and motivated a search for alternative
conditions \cite{MacKenzie,Chen,Wu,Tong2,Jensan,Lidar},
reexamination of some adiabatic algorithms \cite{Wei}, or
generalizations of the adiabatic theorem to open quantum systems
\cite{OpenQS}. While it is valuable to find new conditions that
guarantee adiabaticity in general, it is important to understand why
the traditional adiabatic condition (\ref{Acond}) is sufficient for
some Hamiltonians, but not for others. Moreover, from the practical
point of view it is much easier to work with a simple condition like
(\ref{Acond}) than some other more sophisticated ones. In this
letter, we relate the reported violations of the traditional
adiabatic theorem to resonant transitions between energy levels. We
further show that in the absence of such resonant oscillations, the
traditional adiabatic condition is sufficient to assure
adiabaticity. Our line of thought is close to that of Duki {\em et
al.}~\cite{Duki}, but largely extended with rigorous mathematical
proofs.

It is well known that fast driven oscillations invalidate the
adiabatic theorem \cite{Schiff}. Consider a simple example of a
two-state system driven by an oscillatory force:
      \be
      H(t) = - {1\over 2} \epsilon \sigma_z - V \sin\omega_0 t \
      \sigma_x. \label{Hresonance}
      \ee
We take $V$ to be a small positive number. The exact instantaneous
eigenvalues and eigenstates are
 \be
 E_{0,1} = \mp {1\over 2} \Omega, \qquad
 |E_{0,1}\rangle {=} \left( \begin{array}{c} \alpha^\pm \\
 \pm\alpha^\mp
 \end{array}\right), \quad
 \ee
where $\Omega = \sqrt{\epsilon^2 + 4V^2 \sin^2 \omega_0 t}$ and
$\alpha^\pm = \sqrt{(\Omega \pm \epsilon)/2 \Omega}$. To the lowest
order in $V$, $\Omega \approx \epsilon + (2V^2/\epsilon) \sin^2
\omega_0 t$, $\alpha^+ \approx 1 - (V^2/ 2\epsilon^2) \sin^2
\omega_0 t$, and $\alpha^- \approx (V/\epsilon) \sin \omega_0 t$.
The traditional adiabatic condition (\ref{Acond}) leads to
 \be
 {|\langle E_1|\dot E_0 \rangle| \over E_{10}} \approx {V\omega_0 \over
 \epsilon^2} |\cos \omega_0 t| \ll 1,
 \ee
which is satisfied if $V\omega_0 \ll \epsilon^2$. Near resonance
($\omega_0\approx \epsilon$), this requires $V \ll \epsilon,
\omega_0$. The adiabatic theorem therefore states that if at $t=0$
the system starts in its ground state $|E_0(0)\rangle = (1,0)^T$, it
will stay in the instantaneous ground state at later times. This,
however, is not true as we shall see below.

Using the rotating wave approximation, the wave function of the
system at resonance ($\epsilon = \omega_0$) is given by
 \be
 |\psi (t)\rangle
 = \left( \begin{array}{c} e^{i\omega_0t/2}\cos Vt/2 \\
 e^{-i\omega_0t/2}\sin Vt/2 \end{array}\right).
 \ee
Therefore, the ground state probability
 \be
 P_0(t)=|\langle E_0(t)|\psi(t)\rangle|^2
 \approx (\cos Vt+1)/2 \label{P0}
 \ee
oscillates with the Rabi frequency $f_R=V/2\pi$. At time
$T=T_R/2=\pi/V$, the system will be in the excited state with
probability $P_1=1$, violating the adiabatic theorem. Reducing the
oscillation amplitude $V$ would only increase the Rabi period $T_R$,
and does not keep the system in the ground state. Therefore,
adiabaticity is only satisfied for a time $T \ll T_R$. Indeed some
new versions of adiabatic condition set an upper bound on the time
$T$ in order to guarantee adiabaticity \cite{Wu,Tong2}. However, as
we shall see, this is not necessary in general. Before that, let us
take a close look at the inconsistency discussed in
\cite{Sanders,Tong}.

Let us assume that $H(t)$ is a slowly varying Hamiltonian for which
the adiabatic theorem holds. This means that if at time $t=0$, the
system starts in an eigenstate $|E_n^0\rangle$ ($\equiv
|E_n(0)\rangle$) of $H(0)$, at time $t$, the wave function of the
system will be (see below for proof)
 \be
 |\psi_n(t)\rangle = U(t)|E_n^0\rangle \approx e^{-i\int_0^t
 E_n(t')dt'} |E_n(t)\rangle. \label{psit}
 \ee
Hereafter, we use a gauge in which $\langle E_n(t)|\dot
E_n(t)\rangle = 0$.

Now consider another system with Hamiltonian  (\ref{barH}). The
eigenvalues and eigenstates of the new Hamiltonian are $\bar
E_n(t)=-E_n(t)$ and $|\bar E_n(t)\rangle =
U^\dagger(t)|E_n(t)\rangle$, respectively. From (\ref{psit}), we
have
 \be
 |\bar E_n(t)\rangle \approx e^{i\int_0^t
 E_n(t')dt'} |E_n^0\rangle. \label{En'}
 \ee
It was shown in Refs. \cite{Sanders,Tong} that for system $\bar H$
the adiabatic theorem holds only when (\ref{STcond}) holds, even if
the adiabatic condition (\ref{Acond}) is satisfied. To understand
this, let us write $\bar H$ in the basis $|E_n^0\rangle$:
 \be
 \bar H(t) = \sum_{m,n} \langle E_m^0|\bar H(t)|E_n^0\rangle
 |E_m^0\rangle\langle E_n^0|.
 \ee
However
 \ba
 \langle E_m^0|\bar H(t)|E_n^0\rangle
 &=& -\langle E_m^0|U^\dagger
 H(t)U(t)|E_n^0\rangle \nn
 &=&  -i\langle \psi_m(t)|\dot\psi_n(t)\rangle.
 \ea
Using (\ref{psit}) we find
 \ba
 && \bar H(t) = -\sum_n E_n(t)|E_n^0\rangle\langle E_n^0|
 \label{Hbarnew} \\
 &&-i\sum_{n,m} e^{-i\omega_{nm}(t)}
 \langle E_m(t)|\dot E_n(t)\rangle |E_m^0\rangle\langle E_n^0|.
 \nonumber
 \ea
where $\omega_{nm}(t) \equiv {1\over t}\int_0^t E_{nm}(t')dt'$. The
second line in (\ref{Hbarnew}) involves rapidly oscillating terms
that cause resonant transitions between the levels. The amplitude of
each oscillatory term is $|\langle E_m(t)|\dot E_n(t)\rangle|$.
Hence satisfying (\ref{Acond}) will only reduce this amplitude and,
as we saw before, it does not eliminate Rabi oscillations and
therefore does not keep the system in its original eigenstate beyond
half a Rabi period. Notice that Eq.~(\ref{STcond}) is equivalent to
$|E_n(t) \rangle \approx e^{i\phi(t)}|E_n^0\rangle$, where $\phi(t)$
is some time-dependent phase. This leads to $\langle E_m(t)|\dot
E_n(t)\rangle \propto \langle E_m^0| E_n^0\rangle = 0$. Therefore,
the oscillatory terms in (\ref{Hbarnew}) will all vanish if
(\ref{STcond}) is satisfied, leading to an adiabatic evolution in
agreement with \cite{Sanders,Tong}.

We now provide a general proof for the adiabatic theorem emphasizing
the role of resonant transitions. Let us write the wave function of
the system as:
 \be
 |\psi(t)\rangle = \sum_n a_n(t) e^{-i\int_0^t
 E_n(t')dt'}|E_n(t)\rangle. \label{psisum}
 \ee
For a time-independent Hamiltonian, $a_n(t)$ is a constant while for
a slowly varying Hamiltonian it is a slow function of time.
Substituting (\ref{psisum}) into the Schr\"odinger equation
(\ref{SchEq}), we find
 \ba
 \dot a_m (t) = -\sum_{n\neq m} a_n(t)
 \langle E_m(t)|\dot E_n(t) \rangle e^{-i\int_0^t
 E_{nm}(t')dt'}. \nonumber
 \ea
Integrating over $t$, we get
 \ba
 && a_m(T) - a_m(0) = \label{amT} \\
 &&-\sum_{n\neq m} \int_0^T dt\ a_n(t)
 \langle E_m(t)|\dot E_n(t) \rangle e^{-i\int_0^t
 E_{nm}(t')dt'}. \nonumber
 \ea
To assure adiabaticity, the right hand side of this equation should
be small. With the initial condition $a_m(0) = \delta_{mn}$, this
would immediately yield (\ref{psit}). Since the exponential term in
the integrand of (\ref{amT}) is a rapidly oscillating function, if
the rest of the terms vary very slowly, the integral will be small.
To make this statement more quantitative, let us define the right
hand side of the above equation as the error $\varepsilon_m =
-\sum_{n\neq m} \varepsilon_{nm}$ for the adiabatic evolution, where
 \ba
 \varepsilon_{nm} \equiv \int_0^T dt A_{nm}(t)E_{nm}(t)
 e^{-i\int_0^t E_{nm}(t')dt'}, \label{inteq1}
 \ea
and
 \ba
 A_{nm}(t) \equiv a_n(t)
 {\langle E_m(t)|\dot E_n(t) \rangle \over E_{nm}(t)}. \label{Anm}
 \ea
Using the Fourier transformation:
$ \tilde{A}_{nm}(\omega) = \int_0^T dt e^{i\omega t} A_{nm}(t)$,
we find
 \ba
 \varepsilon_{nm} {=}
 \int {d\omega \over 2\pi} \int_0^T
 dt \tilde{A}_{nm}(\omega)E_{nm}(t) e^{i[\omega -
 \omega_{nm}(t)]t}, \label{inteq2}
 \ea
The integral in (\ref{inteq2}) is suppressed by the oscillatory
exponential in the integrand, except along a path in the two
dimensional $t$-$\omega$ plane defined by the equation $\omega =
\omega_{nm}(t)$, where there is no oscillation. In the presence of
resonant oscillations, $\tilde{A}_{nm}(\omega)$ has finite
components at frequencies $\omega_{nm}(t)$, hence the contribution
from such a dominant path becomes
 \ba
 \varepsilon_{nm} \sim \int_0^T
 dt \tilde{A}_{nm}(\omega_{nm}(t))E_{nm}(t) \nn
 \leq T \max_t |\tilde{A}_{nm}(\omega_{nm}(t))E_{nm}(t)|.
 \ea
The error therefore grows as a function of $T$. As a result, to
assure adiabaticity one needs an upper limit for $T$, as expected
for the case of resonant oscillations. However, this is not the case
for a general system without resonant oscillations, as we shall see
below.

In the absence of such oscillations, $A_{nm}(t)$ can be made as slow
as desired by making the evolution slow. In Fourier space, this
makes $\tilde{A}_{nm}(\omega)$ sharply peaked at low frequencies
with a cutoff frequency $\omega_c$ proportional to the rate of
change of the Hamiltonian. To see this, let us introduce a new
variable $s=t/T$. Since $a_n(t) = a_n(0) + O(\varepsilon_n)$, to the
lowest order in the small error $\varepsilon_n$ we have \cite{note}
 \be
 \tilde{A}_{nm}(\tilde{\omega}) \approx a_n(0) \int_0^1 ds e^{i\tilde{\omega} s}
 {\langle E_m(s)|d/ds|E_n(s) \rangle \over E_{nm}(s)}.
 \ee
where $\tilde{\omega}=\omega T$ is the dimensionless frequency.  The
integral on the right hand side is independent of $T$. Let
$\tilde{\omega}_c$ be the largest dimensionless frequency of
$\tilde{A}_{nm}(\tilde{\omega})$. Therefore $\omega_c =
\tilde{\omega}_c/T$ can be made arbitrarily small by making $T$
large. Notice that $\tilde{\omega}_c$ is a constant that only
depends on the properties of the Hamiltonian and does not depend on
the evolution time $T$.

If $\omega_c \ll \omega_{nm}(t)$, one can neglect $\omega$ in the
exponential in the integrand of (\ref{inteq2}) and therefore perform
the $t$- and $\omega$-integrations independently, yielding
 \ba
 \varepsilon_{nm} &\sim& \omega_c |\tilde{A}_{nm}(0)|
 \leq \omega_c \int_0^T |A_{nm}(t)|dt \nn
 &\leq& \omega_c T \max_t |A_{nm}(t)| \nn
 &\leq& \tilde{\omega}_c \max_t \left|{\langle E_m(t)|\dot E_n(t) \rangle
 \over E_{nm}(t)} \right|.
 \ea
Therefore, $\varepsilon_{nm}$ can be made arbitrarily small by only
satisfying the adiabatic condition (\ref{Acond}).

The same conclusion can also be reached from a different angle.
Using integration by parts, Eq.~(\ref{inteq1}) becomes
 \ba
 \varepsilon_{nm} = \left[A_{nm}(T) e^{-i\int_0^T
 E_{nm}(t')dt'}-A_{nm}(0)\right] \nn
  - \int_0^T dt \dot A_{nm}(t)e^{-i\int_0^t E_{nm}(t')dt'} \nn
 \leq |A_{nm}(T)| + |A_{nm}(0)| + \int_0^T dt |\dot A_{nm}(t)|.
 \ea
The last term above is responsible for the breakdown of the
adiabatic theorem. Let us define $t_i$, $i=1,...,M_{nm}$, as the
solutions to $\dot A_{nm}(t_i)=0$, where $M_{nm}$ is the number of
zeros of $\dot A_{nm}(t)$ in the interval $[0,T]$. Since $\dot
A_{nm}(t)$ is monotonic between $t_i$ and $t_{i+1}$, we can write
 \ba
 \int_0^T dt |\dot A_{nm}(t)| &=&
 \sum_{i=0}^{M_{nm}} \left|\int_{t_1}^{t_{i+1}} dt \dot A_{nm}(t)\right| \nn
 &=& \sum_{i=0}^{M_{nm}} \left|A_{nm}(t_{i+1})-A_{nm}(t_i)\right|,
 \ea
where we have defined $t_0=0$ and $t_{M_{nm}+1}=T$. Thus
 \ba
 \varepsilon_{nm} &\leq& 2\sum_{i=0}^{M_{nm}} \left|A_{nm}(t_i)\right|
  \leq 2M_{nm} \max_{t\in [0,T]} \left|A_{nm}(t)\right| \nn
  &\leq& 2M_{nm} \max_{t\in [0,T]}
 \left|{\langle E_m(t)|\dot E_n(t) \rangle \over E_{nm}(t)}\right|.
 \label{enm}
 \ea
Since the error depends on $M_{nm}$, it is important to understand
how $M_{nm}$ depends on the evolution time $T$.

Let us first consider a Hamiltonian that has an oscillatory term
with frequency $\omega_0$. Oscillations of the Hamiltonian will
create oscillations in $A_{nm}(t)$ and therefore the number of zeros
of $\dot A_{nm}(t)$ will grow with time as $M_{nm} \sim \omega_0T$.
In that case, without putting an upper bound on $T$, it is not
possible to limit the error $\varepsilon_{nm}$. This is in agreement
with our previous observation for cases involving resonant
transitions, as well as the additional conditions introduced in
Refs.~\cite{Wu,Tong2}. On the other hand, if $M_{nm}$ does not grow
with time, one can always reduce $\varepsilon_{nm}$ by just
satisfying (\ref{Acond}) without a need to limit $T$. To see this,
let us again use the dimensionless parameter $s = t/T$. If by
slowing down the evolution, we only change $T$ and not other
parameters in the Hamiltonian, then the Hamiltonian $H(s)$ and its
eigenvalue $E_n(s)$ and eigenfunctions $|E_n(s)\rangle$ will all be
independent of $T$. Again to the lowest order in $\varepsilon_n$,
$a_n(t)\approx a_n(0)$ and therefore from (\ref{Anm}), $A_{nm}(t)
\approx a_n(0) \langle E_m(t)|\dot E_n(t) \rangle / E_{nm}(t)$
\cite{note}. The times $t_i$ can therefore be obtained by solving
 \be
 \dot A_{nm}(t) = {a_n(0)\over T^2} {d \over ds} {\langle E_m(s)|d/ds|
 E_0(s) \rangle \over E_{nm}(s)} = 0.
 \ee
The number of zeros of this equation, i.e. $M_{nm}$, is therefore
finite and independent of $T$. In that case, (\ref{enm}) assures
that by just satisfying the adiabatic condition (\ref{Acond}), the
error $\varepsilon_{nm}$ can be made as small as desired.

From the above proof it becomes evident that the following way of
stating the adiabatic theorem removes all the inconsistences: For a
Hamiltonian $H(s)$, where $s = t/T \in [0,1]$, the evolution of the
system starting from an eigenstate $|E_n(0)\rangle$ is adiabatic
provided that
 \be
 T \gg \max_{s\in [0,1]}
 {|\langle E_m|dH/ds|E_n \rangle| \over E_{nm}^2} \qquad \text{for all } m\neq
 n. \label{NewAT}
 \ee

It should be emphasized that our goal in this paper was just to
study the sufficiency of the adiabatic condition (\ref{Acond}) for
adiabatic evolution without worrying about the scaling issue. In
other words, we do not discuss dependence of the error
$\varepsilon_{nm}$ on the system size. Scaling becomes important for
determining the performance of an adiabatic quantum computer. The
exponentially large number of states involved in the sum in
(\ref{amT}) requires $\varepsilon_{nm}$ to be exponentially small in
order to keep the sum small. Fortunately, this does not put a
stringent limitation on the time $T$. To understand this, let us
write the error as
 \be
 \varepsilon_{nm} \lesssim {1\over T} \max_{s\in [0,1]}
 {|\langle E_m|dH/ds|E_n \rangle| \over E_{nm}^2}.
 \ee
Typically $\langle E_m|dH/ds|E_n \rangle$ is exponentially small
otherwise the curvature of the energy levels
 \ba
 {d^2 E_n \over ds^2} = 2\sum_{m \ne n}{|\langle E_m|dH/ds|E_n \rangle|^2 \over
 E_{nm}} + \langle E_n|{d^2H\over ds^2}|E_n \rangle \nonumber
 \ea
becomes extremely large due to the sum over exponentially large
number of terms. For the simple example of adiabatic Grover search
\cite{Roland}, it is easy to show that $\langle E_m|\dot H|E_0
\rangle = 0$ for $m>1$, therefore only first two energy levels
contribute to the adiabatic evolution. For problems with local
interactions, the matrix elements typically decay exponentially with
the energy separation between the states. This can be checked
perturbatively near the beginning and the end of the evolution
(using similar techniques as in Ref.~\cite{localminima}). It can
also be tested numerically for systems with not very large size
\cite{Jordan}. Such exponential suppression of the matrix elements
allows only a few energy levels to participate in the calculation of
the error. Especially, in adiabatic quantum computation when the gap
between the ground state and the first excited state becomes much
smaller than other energy separations, those two states dominantly
determine the error of the computation and the evolution time can be
determined by the minimum gap between those, as has been confirmed
numerically for up to 20 qubits \cite{Amin}. More work is needed to
make these statements mathematically rigorous. Moreover, a realistic
adiabatic quantum computer will always couple to an environment.
Therefore, other methods \cite{OpenQS,Amin} are necessary to study
the evolution of such open quantum systems.

To summarize, we have shown that the inconsistencies in the
traditional adiabatic theorem reported in the literature are all
closely related to the fact that for systems subject to fast driven
oscillations, resonant transitions between energy levels cannot be
suppressed by just reducing the amplitude of oscillations, although
the adiabatic condition (\ref{Acond}) can be satisfied. Since the
amplitude of oscillations determine the Rabi frequency, reducing the
amplitude would only increase the Rabi period. Within a time of the
order of half a Rabi period, the system will undergo a transition
out of its original state. Thus, the Rabi period sets an upper limit
for the total time of the adiabatic evolution. On the other hand, if
the Hamiltonian of the system does not involve any driven
oscillations, there is no such mechanism to take the system out of
its original state and the traditional adiabatic condition is
adequate to guarantee adiabaticity.

The author thanks D.V.~Averin, A.J.~Berkley, F.~Brito, V.~Choi,
R.~Harris, J.~Johansson, M.~Johnson, W.M.~Kaminsky, T.~Lanting,
D.~Lidar, R.~Raussendorf, and G.~Rose for fruitful discussions and
E.~Farhi for suggesting (\ref{NewAT}) as a consistent way of
expressing the adiabatic theorem.


\begin{thebibliography}{99}
\bibitem{note} The approximation becomes self-consistent by
making sure that the error $\varepsilon_n$ is small at the end.
\end{thebibliography}
\end{document}